\documentclass[reprint, aps, prb, amsmath, amssymb, superscriptaddress, floatfix, longbibliography]{revtex4-2}
\usepackage{graphicx}
\usepackage{xcolor}
\usepackage{bm}
\usepackage{soul}
\usepackage[normalem]{ulem}
\usepackage[hypertexnames=false,colorlinks=true,citecolor=blue,linkcolor=blue,urlcolor=blue]{hyperref}
\usepackage{comment}
\usepackage{physics}
\usepackage[T1]{fontenc}
\usepackage[utf8]{inputenc}
\usepackage{pifont}
\usepackage{cancel}
\usepackage{wrapfig}
\usepackage[stable]{footmisc}
\usepackage{array}
\usepackage{gensymb}
\usepackage{multirow}
\usepackage{hhline}
\newcommand{\cmark}{\ding{51}}%
\newcommand{\xmark}{\ding{55}}%
\usepackage{chemformula}
\def\nn{\nonumber}
\def\bea{\begin{eqnarray}}
\def\eea{\end{eqnarray}}
\def\be{\begin{equation}}
\def\ee{\end{equation}}

\def\kb{{\bm k}}

\def\bal{\begin{aligned}}
\def\eal{\end{aligned}}
\begin{document}

\title{Asymmetric Scattering-Induced N\'eel Spin-Orbit Torque in Antiferromagnets}

\author{Sayan Sarkar}
\email{sayans21@iitk.ac.in}
\author{Amit Agarwal}
\email{amitag@iitk.ac.in}
\affiliation{Department of Physics, Indian Institute of Technology Kanpur, Kanpur 208016}
\begin{abstract}
    Magnetic switching in antiferromagnets relies on N\'eel spin orbit torque (NSOT), which originates from a current-induced staggered spin polarization of itinerant electrons. In collinear antiferromagnets, such a response requires the spin susceptibility to be odd under combined space–time inversion symmetry ($\cal PT$), and is conventionally attributed to symmetric scattering processes. Here, we demonstrate that asymmetric impurity scattering generates an additional $\cal PT$-odd spin polarization when coupled with the anomalous spin polarizability (ASP) of Bloch electrons. This extrinsic contribution arises from the interplay between antisymmetric higher-order scattering processes and band geometry, effectively converting an otherwise $\cal PT$-even susceptibility into a staggered spin polarization. Using a minimal model of tetragonal \ch{CuMnAs}, we show that this anomalous skew-scattering contribution can be comparable to, and with sufficient impurity density even exceed, the conventional symmetric scattering (Drude) contribution. Our results identify a new band-geometry-driven mechanism for NSOT and establish an efficient route for electrical control of antiferromagnets.
\end{abstract}

\maketitle

\section{Introduction}

Magnetic memory devices are a central component of modern spintronic technologies due to their non-volatility, scalability, and energy-efficient operation~\cite{Tsymbal_19,stuart_science2008, Miron_nature2011, Liu_Science2012, Wang_JOPD2013, Kent_NN2015, Wong_nature2015,Difalco_inorganics2025,Cai_NSR2023}. Conventional implementations rely on ferromagnets (FMs), in which information is encoded in the orientation of a finite magnetization that can be switched by current-induced spin torques. However, the presence of stray magnetic fields in FMs leads to undesirable cross-talk between neighboring bits and limits device miniaturization~\cite{Cai_NSR2023,Shao_IEEE2021,Wu_2020,Bernard_IEEE2024,Baek_pra2019,sato_2018,Bhatti_MT2017,Jenkins_JOPD2020}. In this context, antiferromagnets (AFMs), characterized by vanishing net magnetization due to antiparallel alignment of magnetic moments, have emerged as a promising alternative. The absence of stray fields allows for dense device integration, while their intrinsically faster spin dynamics (typically three orders of magnitude higher than in FMs) enable ultrafast operation, making AFMs highly attractive for next-generation memory applications~\cite{Chengbao_AEM20219,Zelenzy_2017,Yutaro_SA2025,Xiong_FR2022}.

\begin{figure}[tb]
    \centering
    \includegraphics[width=1\linewidth]{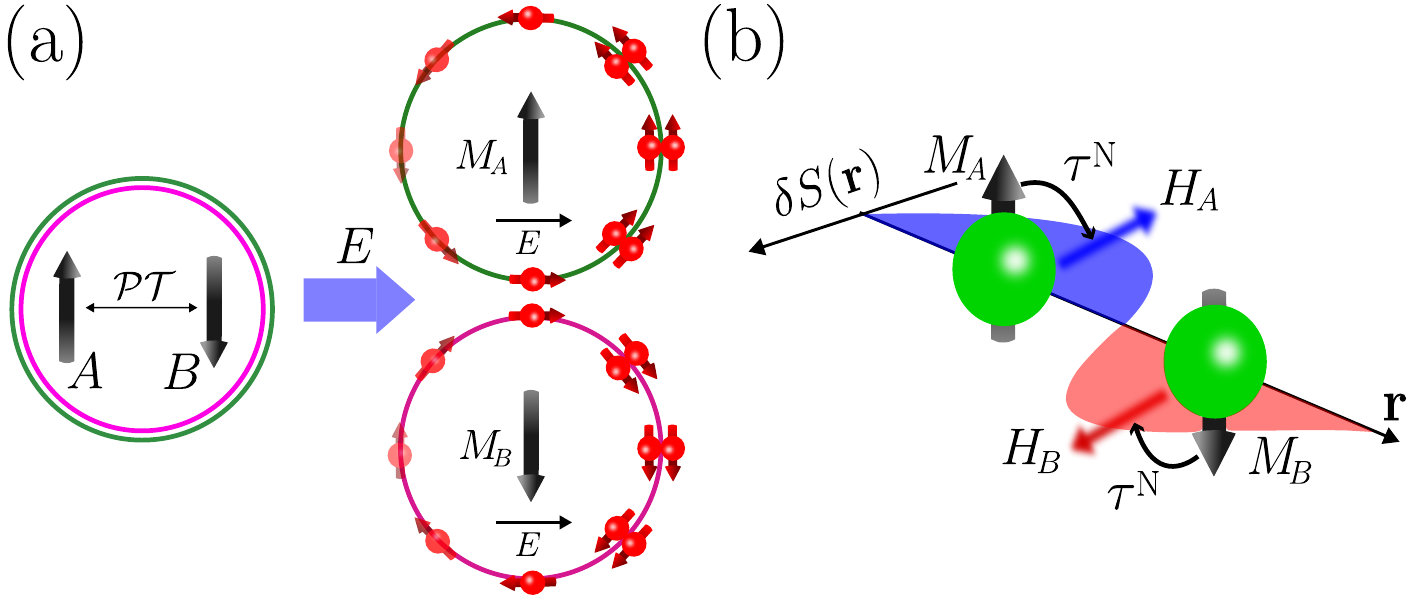}
    \caption{\textbf{Schematic of N\'eel spin polarization.} 
    (a) $\cal PT$-related Fermi surfaces of a collinear antiferromagnet with opposite magnetic sublattices ($A$ and $B$). Under an applied electric field $\bm E$, itinerant electrons generate a staggered spin polarization with opposite signs on the two sublattices. 
    (b) These sublattice-resolved spin polarizations act as effective fields ($\bm H_A$ and $\bm H_B$), producing a N\'eel spin–orbit torque ($\tau^{\rm N}$) that enables electrical switching of the N\'eel vector.}
    \label{fig1}
\end{figure}

A fundamental challenge in AFM spintronics is the efficient electrical manipulation of the N\'eel order parameter. This challenge has been addressed through N\'eel spin-orbit torque (NSOT), in which an applied electric field generates a staggered spin polarization that acts oppositely on the two magnetic sublattices. This exerts a torque on the N\'eel vector~\cite{Wadley_SA2016,Bodnar_NC2018,Tristan_pra2018,Chen_NM2019,zelenzy_prl2014,Zelenzy_prb2017,Tang_NC2025,sarkar_Neel2025}. In collinear AFMs preserving combined space--time inversion ( $\cal PT$) symmetry, the two sublattices are related by the $\cal PT$ operation. This enforces that only $\cal PT$-odd spin responses can produce such staggered spin polarization~\cite{zelenzy_prl2014,Zelenzy_2017}, as schematically demonstrated in Fig.~\ref{fig1}. Microscopically, current-induced spin polarization arises from non-equilibrium carrier dynamics in the presence of an electric field and disorder. Within existing theoretical frameworks, the dominant contribution relevant for NSOT originates from symmetric scattering processes~\cite{Edelstein_SSC1990,Thole_JAP2020, Ingrid_prr2021,Johansson_2024}. Their associated spin susceptibility is $\cal PT$-odd, so the net spin polarization vanishes, while compensating local spin polarization remains finite on each sublattice. In contrast, the interband-coherence contribution governed by the anomalous spin polarizability (ASP) is $\cal PT$-even. Therefore, it cannot by itself generate the staggered spin polarization required for NSOT.

Here, we identify a previously overlooked mechanism for generating N\'eel spin polarization in $\cal PT$-symmetric AFMs, arising from asymmetric impurity scattering~\cite{Guo_prb2024,xiao_PRB2017_semi,Xiao_prb2019,Song_prl2023}. We show that the antisymmetric component of higher-order scattering process, intrinsically linked to the Berry curvature of the electronic bands, combines with the anomalous spin polarizability to produce a new $\cal PT$-odd spin susceptibility. This provides an additional extrinsic route to NSOT beyond the conventional Drude mechanism. We explicitly demonstrate this using a minimal model of tetragonal CuMnAs. We calculate the staggered spin response and show that the anomalous skew-scattering channel can compete with, and in a moderate disorder regime, even exceed the Drude contribution. We further use the resulting torque scale for an illustrative switching analysis. Our work establishes a band-geometry-driven disorder mechanism for electrical control of antiferromagnets and clarifies the parameter regime in which it becomes experimentally relevant.

\section{Asymmetric scattering contribution to spin magnetization}

In the semiclassical Boltzmann framework, current-induced spin polarization arises from corrections to both the spin expectation value and the non-equilibrium distribution function in the presence of an external electric field $\bm{E}$ and weak disorder. These corrections can be systematically organized according to their order in the electric field and impurity potential. The total spin polarization is obtained by combining all allowed contributions from these corrections. In this section, we identify a previously overlooked contribution originating from the interplay between disorder-induced spin correction and asymmetric impurity scattering. The total spin polarization is given by
\be\label{spin_def}
\delta S^{\nu} = \sum_l s^{\nu}_l f_l~,
\ee
where $\nu$ denotes the spin polarization component. The summation runs over Bloch states $\ket{l}=\ket{n {\bm k}}$ with $n$ and ${\bm k}$ being the band index and crystal momentum, respectively. Here, $f_l$ is the non-equilibrium distribution function, and $s^{\nu}_l$ is the expectation value of the spin operator in the presence of perturbations and impurity. The spin expectation value, including corrections to the Bloch state induced by the external electric field ($\bm{E}$)~\cite{Edelstein_SSC1990,Thole_JAP2020, Ingrid_prr2021,Johansson_2024,Sarkar_njp2025,zhang_PRB2024_int,Ye_prb2024} and impurity scattering~\cite{Guo_prb2024,xiao_PRB2017_semi,Xiao_prb2019,Du_NC2019,varshney_2026,varshney_2026_longitude}, can be naturally decomposed as
\be\label{total_spin}
s^{\nu}_l = s_l^{\nu,0} + s_l^{\nu,{\rm anm}} + s_l^{\nu,{\rm sct}},
\ee
where $s_{l}^{\nu,0}=\langle {n\bm{k}}|\hat{s}^\nu|{n\bm{k}}\rangle=\langle u_{n\bm{k}}|\hat{s}^\nu|u_{n\bm{k}}\rangle$ is the spin expectation in the unperturbed Bloch state, where $|u_{n\mathbf{k}}\rangle$ denotes the cell-periodic part of $\ket{n\bm k}$. The anomalous term $s_{l}^{\nu,{\rm anm}}$ arises from electric-field-induced interband coherence and is given by
\be
s_{l}^{\nu,{\rm anm}}=-\frac{e}{\hbar}\gamma_{n\bm{k}}^{\nu,b}E_b.
\ee
Here, $\gamma^{\nu,b}_{n\bm{k}}$ is the anomalous spin polarizability (ASP)~\cite{xiao_prl2023,Guo_prb2024,sarkar_mtq2026} and defined as 
\be\label{eq:SBC}
\gamma^{\nu,b}_{n\bm{k}}
=-2\hbar^{2} \mathrm{Im} \sum_{n' \ne n}
\frac{s^\nu_{nn'}({\bm k}) v^b_{n'n}({\bm k})}
{(\varepsilon_{n {\bm k}} - \varepsilon_{n' {\bm k}})^{2}}~.
\ee
In this equation, for a given operator $\hat{A}$ we define
$A_{n'n}(\mathbf{k}) \equiv \langle u_{n' \mathbf{k}} | \hat{A} | u_{n \mathbf{k}} \rangle$ and $\varepsilon_{n\bm{k}}$ denotes the band dispersion for the Bloch-state $|n\bm{k}\rangle$. This term captures the band-geometric origin of the intrinsic current-induced spin polarization~\cite{Shen_prl2014,zelenzy_prl2014,Guo_prb2024,xiao_prl2022,sanjay_ESSC2026,Fregoso_prb2022}.

\begin{figure}[tb]
    \centering
    \includegraphics[width=1\linewidth]{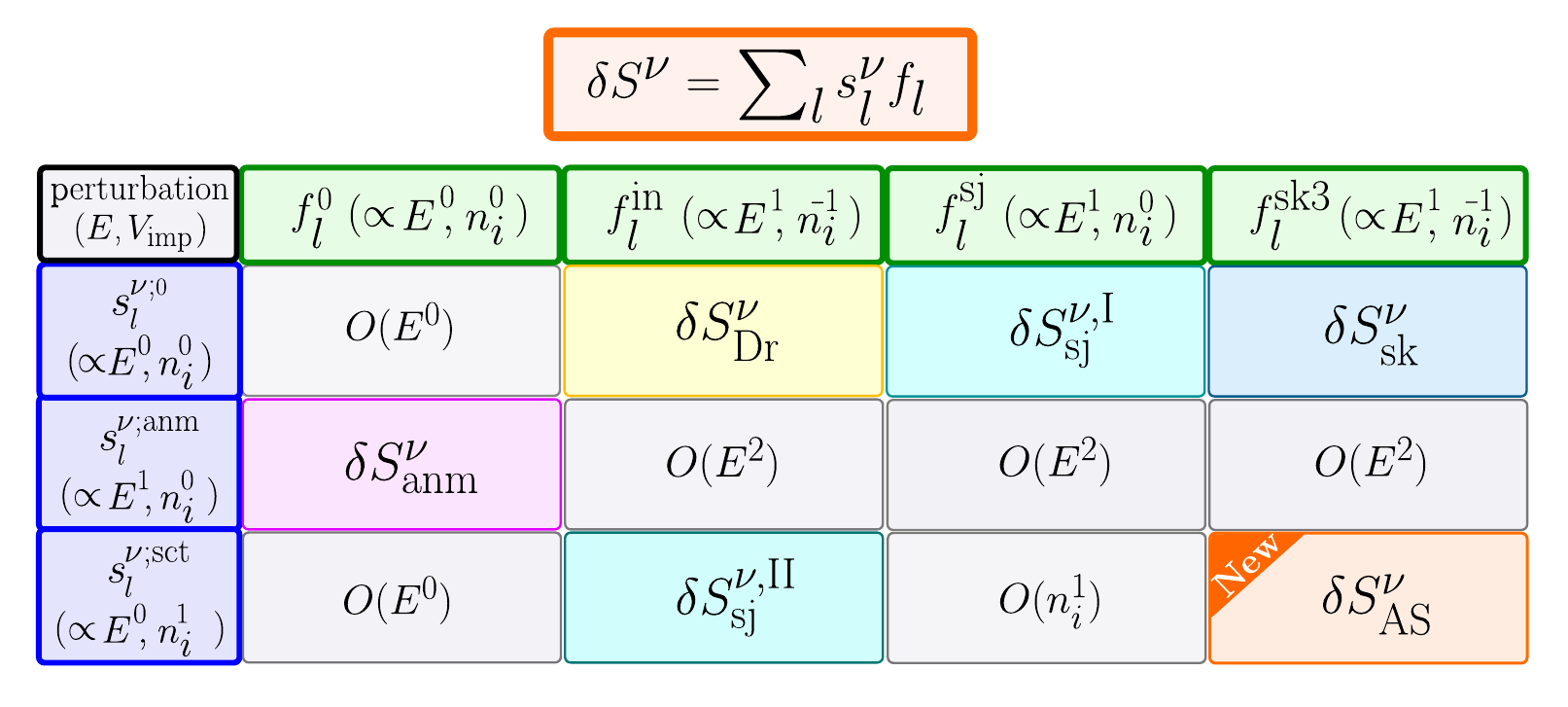}
    \caption{\textbf{Schematic illustration of contributions to current-induced spin polarization.} 
    An applied electric field ($\bm E$) and impurity potential ($V_{\rm imp}$) induce corrections to both the spin expectation value ($s_l^\nu$) and the distribution function ($f_l$). The interplay of these corrections gives rise to five distinct contributions to the linear-order spin polarization ($\propto E$), retained up to zeroth order in impurity density ($n_i^0$). 
    In the weak-disorder limit, the relaxation time scales as $\tau \propto n_i^{-1}$, and higher-order impurity-induced contributions are neglected.}
    \label{fig2}
\end{figure}

Disorder induced asymmetric scattering further modifies the spin expectation value through a correction $s^{\nu,{\rm sct}}_l$, which represents the spin analogue of the side-jump velocity~\cite{xiao_PRB2017_semi, Xiao_prb2019}. For short-range delta function-like impurity potentials $V_{\rm imp}(\bm{r})=\sum_i V_i\delta(\bm{r}-R_i)$, it takes the form
\bea\label{sjv} 
s^{\nu,{\rm sct}}_{l} &=& -2\pi \sum_{n' {\bm k}'} W_{{\bm k}{\bm k}'} \delta(\varepsilon_{n {\bm k}} - \varepsilon_{n' {\bm k}'}) \nn\\ 
&\times& \mathrm{Im}\Bigg[ 
\sum_{n'' \neq n'} 
\frac{ \langle u_{n {\bm k}} | u_{n' {\bm k}'} \rangle 
\langle u_{n'' {\bm k}'} | u_{n {\bm k}} \rangle 
s^\nu_{n' n''}({\bm k'})} 
{\varepsilon_{n' {\bm k}'} - \varepsilon_{n'' {\bm k}'}} \nn\\ 
&-& 
\sum_{n'' \neq n} 
\frac{\langle u_{n'' {\bm k}} | u_{n' {\bm k}'} \rangle 
\langle u_{n' {\bm k}'} | u_{n {\bm k}} \rangle 
s^\nu_{n n''}({\bm k})} 
{\varepsilon_{n {\bm k}} - \varepsilon_{n'' {\bm k}}} 
\Bigg], 
\eea
where $W_{{\bm k}{\bm k}'} = \langle |V^{0}_{{\bm k}{\bm k}'}|^{2} \rangle_{\rm dis}$ is the disorder-averaged Born amplitude for impurity scattering. Here, $\langle \cdots \rangle_{\rm dis}$ denotes configurational averaging over random impurity distributions. Assuming that impurity scattering does not induce inter-band transitions in the DC response (i.e. $n'=n$), the scattering rate is peaked near $\kb'\to\kb$ (see Eq.~\eqref{ScRt2}). In this limit, Eq.~\eqref{sjv} simplifies to~\cite{Guo_prb2024},
\be\label{Mod_sct}
s^{\nu,\mathrm{sct}}_l \simeq \dfrac{\gamma^{\nu,b}_{n\bm{k}} k_b}{\tau},
\ee
where $\tau$ is the momentum relaxation time. In the present manuscript, this compact form is obtained in the weak disorder limit. The detailed derivation is provided in Appendix~\ref{SS_correction}.

The non-equilibrium distribution function can also be expanded in powers of the perturbation field and impurity potential as follows,
\be\label{fermi_cont}
f_l = f_{n\bm{k}}^0 + f_{n\bm{k}}^{\mathrm{in}} + f_{n\bm{k}}^{\mathrm{sj}} + f_{n\bm{k}}^{\mathrm{sk3}}. 
\ee
The leading order corrections in field ($\bm{E}$) and impurity density ($n_i$) are 
\bea
f_{n\bm{k}}^{\rm in}&=&\dfrac{e\tau}{\hbar}E_b\partial_{k_b} f_{n\bm{k}}^0,\\
f_{n\bm{k}}^{\rm sj}&=&\tau \sum_{\bm{k}'} w^{(2),{\rm cs}}_{n;\bm{k}'\bm{k}}\bigl(f_{n\bm{k}}^0 - f_{n\bm{k}'}^0\bigr),\\
f_{n\bm{k}}^{\rm sk3}&=&-\frac{e\tau^2}{\hbar}\sum_{\bm{k}'} w^{(3),a}_{n;\bm{k}'\bm{k}}E_b\left(\partial_{k_b} f_{n\bm{k}}^0+\partial_{k'_b} f_{n\bm{k}'}^0\right).
\eea
Here, $w^{(n)}$ denotes the $n$-th order scattering rate. $w^{(2)}$ is by construction symmetric under $\kb \leftrightarrow\kb'$ ($w^{(2)}_{n;\bm{k}'\bm{k}}=w^{(2)}_{n;\bm{k}\bm{k}'}$). On the other hand, $w^{(3),a}$ is the antisymmetric part of $w^{(3)}$ ($w^{(3),a}_{n;\bm{k}'\bm{k}}=-w^{(3),a}_{n;\bm{k}\bm{k}'}$) responsible for skew scattering. In the present disorder model it is linked to the band-resolved Berry curvature, which supplies the symmetry structure needed below to convert the $\mathcal{PT}$-even ASP into a $\mathcal{PT}$-odd response channel~\cite{Du_NC2019,varshney_2026,sanjay_ESSC2026}. It is given by, 
\bea\label{sk3_R}
&&w^{(3),a}_{n;\bm{k}\bm{k}'}= \frac{2\pi^2 n_i V_1^3}{\hbar}\sum_{ \kb''}\big[(\kb'\times\kb'')+(\kb\times\kb')\nn\\&&+(\kb''\times\kb)\big].\bm{\Omega}_ n(\bm{k}) \delta(\varepsilon_{ n \bm k}- \varepsilon_{ n \bm k'})\delta(\varepsilon_{ n\bm k} - \varepsilon_{ n \bm k''}),~~~~~~
\eea
where $V_1$ is the first moment of the impurity potential (see Eq.~\eqref{sct3_pot}). Please refer to Appendices~\ref{Sct_cor_func} and~\ref{Sct_rate} for detailed derivations of the corrections to the non-equilibrium distribution function and the scattering rates.

\begin{table}[tbp]
    \centering
    \caption{\textcolor{black}{Symmetry classification of the linear-order spin-susceptibility contributions generated by the electric-field perturbation and impurity scattering. The \cmark\ (\xmark) denotes an even (odd) response under the corresponding symmetry. The starred susceptibility component is the new $\cal PT$-odd response discussed in this work; see Eq.~\eqref{New_Spin}.}}
    \renewcommand{\arraystretch}{1.5}
    \setlength\tabcolsep{0.3cm}
    \begin{tabular}{c c c c c}
    \hline
    \hline
    Spin Susceptibilities & $\mathcal{P}$ & $\mathcal{T}$ & $\mathcal{PT}$ &  Refs. \\
    \hline
    $\chi^{\nu;b}_{\mathrm{Dr}}$ & \xmark & \cmark & \xmark & \cite{Ingrid_prr2021,Johansson_2024,Tang_NC2025,Ye_prb2024,Thole_JAP2020} \\
    $\chi^{\nu;b}_{\mathrm{anm}}$, $\chi^{\nu;b}_{\mathrm{sj}}$, $\chi^{\nu;b}_{\mathrm{sk}}$ & \xmark & \xmark & \cmark & \cite{Lai_prl2025,Guo_prb2024,xiao_PRB2017_semi,Xiao_prb2019} \\
    $\chi^{\nu;b}_{\mathrm{AS}}*$ & \xmark & \cmark & \xmark & Current work \\
    \hline
    \hline
    \end{tabular}
    \label{Table1}
\end{table}
    

Combining all of these corrections to both spin expectation and the non-equilibrium distribution yields the standard contributions to the linear spin Edelstein effect (LSEE), including the Drude ($\delta S^{\nu}_{\rm Dr}$), anomalous ($\delta S^\nu_{\rm anm}$), side-jump ($\delta S^\nu_{\rm sj}$), and skew-scattering ($\delta S^\nu_{\rm sk}$) terms. For clarity, we use `Dr' for the conventional symmetric intraband contribution, `anm' for the intrinsic ASP-driven interband contribution, `sj' for the side-jump-like disorder correction and `sk' for the conventional skew-scattering term. Fig.~\ref{fig2} summarizes these channels and their leading impurity-density scaling. These are the standard spin polarization components widely discussed in the literature~\cite{Guo_prb2024,xiao_PRB2017_semi,Xiao_prb2019}. The corresponding spin susceptibility components $\chi^{\nu;b}_i$ can then be obtained from the constitutive relation
\be
\delta S^\nu_{i}=\chi^{\nu;b}_iE_b~~~~(i=\rm Dr,anm,sj,sk).
\ee
Among all the above contributions to the spin susceptibility, only $\chi^{\nu;b}_{\mathrm{Dr}}$ is a $\cal PT$-odd quantity, while the remaining components ($\chi^{\nu;b}_{\mathrm{anm}},\chi^{\nu;b}_{\mathrm{sj}}$ and $\chi^{\nu;b}_{\mathrm{sk}}$) are $\cal PT$-even~(see Table~\ref{Table1}). At linear order in the applied electric field, the standard channels can be organized as follows: $s^{\nu,0}f^{\rm in}$ gives the Drude term, $s^{\nu,\rm anm}f^0$ gives the intrinsic anomalous term, $(s^{\nu,\rm sct}f^{\rm in}+s^{\nu,0}f^{\rm sj})$ forms the side-jump class, and $s^{\nu,0}f^{\rm sk3}$ gives the conventional skew-scattering response. The product $s^{\nu,\rm anm}f^{\rm sk3}$ is second order in the electric field and is therefore beyond the present linear-response theory, while $s^{\nu,\rm sct}f^{\rm sj}$ is found in the higher-order impurity-density sector ($\propto n_i^1$) and therefore neglected. 

In addition to these known contributions, we identify a previously overlooked term arising from the cross-combination of the disorder-induced spin correction $s^{\nu,\mathrm{sct}}_{n\bm{k}}$ and the asymmetric distribution function $f^{\mathrm{sk3}}_{n\bm{k}}$. This contribution is given by
\bea\label{New_Spin}
\delta S^{\nu}_{\rm AS}
&=&\sum_{n\bm{k}}s^{\nu,\mathrm{sct}}_{n\bm{k}}f^{\mathrm{sk3}}_{n\bm{k}},\\
&=& \frac{e\tau}{\hbar}\sum_{n\bm{k}\bm{k}'}w^{(3),a}_{n;\bm{k}\bm{k}'}{\gamma^{\nu,c}_{n\bm{k}} k_c} \left(\partial_{k_b} f_{n\bm{k}}^0+\partial_{k_b'} f_{n\bm{k}'}^0\right)E_b.\nn
\eea
The suffix `AS' denotes anomalous skew-scattering, following the terminology introduced in Ref.~\cite{Song_prl2023}. This term is distinct from conventional skew-scattering contributions because it is a cross term. It couples the asymmetric scattering rate $w^{(3),a}$ with the anomalous spin polarizability $\gamma^{\nu,c}_{n\bm{k}}$, rather than with the uncorrected spin expectation value alone. Since $w^{(3),a}$ is $\mathcal{PT}$-odd~\cite{Song_prl2023} while $\gamma^{\nu,c}_{n\bm{k}}$ is $\mathcal{PT}$-even~\cite{Guo_prb2024,xiao_prl2022,Sarkar_njp2025}, their product yields a $\mathcal{PT}$-odd spin susceptibility ($\chi^{\nu;b}_{\mathrm{AS}}$). In physical terms, asymmetric scattering biases carriers in momentum space in a Berry-curvature-dependent way, converting the otherwise $\cal PT$-even ASP into a $\cal PT$-odd observable. The ${\cal P}{\cal T}$ character of the response has a direct sublattice consequence. For the two magnetic sublattices ($A$ and $B$) related by ${\cal P}{\cal T}$, the calculated ${\cal P}{\cal T}$-odd susceptibilities satisfy $\chi^{\nu;b}_{\mathrm{o},A}=-\chi^{\nu;b}_{\mathrm{o},B}$, whereas ${\cal P}{\cal T}$-even channels satisfy $\chi^{\nu;b}_{\mathrm{e},A}=\chi^{\nu;b}_{\mathrm{e},B}$ and therefore do not generate a N\'eel torque. This is why the `AS' term is relevant for NSOT even though it is built from an ASP ingredient that is ${\cal P}{\cal T}$-even by itself.


The remainder of the manuscript is organized accordingly. In Sec.~III we evaluate the ${\cal P}{\cal T}$-odd spin-susceptibility channels in tetragonal CuMnAs and verify their staggered sublattice structure. In Sec.~IV we isolate the disorder prefactor that controls when the anomalous skew scattering term can compete with the Drude channel, and in Sec.~V we use the resulting torque scale in an illustrative switching calculation.

\section{N\'eel spin polarization in $\ch{CuMnAs}$}
The magnetic point groups of collinear AFMs such as \ch{MnBi2Te4}, \ch{CuMnAs}, \ch{Mn2Au} possess combined $\cal PT$ symmetry~\cite{Wadley_SA2016,Bodnar_NC2018}. In these systems, the two opposite magnetic sublattices are related by the $\cal PT$ operation. For a bipartite collinear AFM with sublattices $A$ and $B$, the symmetry transformation acts as $\mathcal{PT}: \mathbf{r}_A \to \mathbf{r}_B$ and $\mathcal{PT}: \mathbf{M}(\mathbf{r}_A) \to -\mathbf{M}(\mathbf{r}_B)$, where $\mathbf{M}(\mathbf{r})$ denotes the local magnetic moment. This relation ensures the compensation of the net atomic magnetization in equilibrium.

In the presence of $\cal PT$ symmetry, response tensors can be classified according to their transformation properties. In particular, for a $\cal PT$-odd spin susceptibility, the symmetry operation enforces $\mathcal{PT}: \chi^{\nu;b}_{A} \to -\chi^{\nu;b}_{B}$. As a consequence, the induced itinerant spin polarization on the two sublattices acquires opposite signs, resulting in a staggered spin response. Such a staggered spin polarization acts as the driving mechanism for the NSOT and enables electrical switching in $\cal PT$-symmetric AFMs. In the sublattice-resolved numerical results below, this distinction is realized explicitly. The Drude and anomalous skew-scattering susceptibilities are odd under sublattice exchange, while the ${\cal P}{\cal T}$-even channels remain equal on the two sublattices and therefore drop out of the N\'eel response.

For a numerical demonstration of the above mechanism, we consider a minimal tight-binding model of tetragonal \ch{CuMnAs} on a crinkled quasi-two-dimensional square lattice with collinear antiferromagnetic order~\cite{Jungwirth_prl2017},
\bea\label{Ham_cumnas}
   \mathcal{H}&=&-2t\cos{\frac{k_x}{2}}\cos{\frac{k_y}{2}}\tau_x\sigma_0-t'(\cos{k_x}+\cos{k_y})\tau_0\sigma_0\nn\\&&+\lambda \tau_z(\sigma_y \sin{k_x}-\sigma_x \sin{k_y})+J_n\tau_z \mathbf{\sigma} \cdot \hat{\bm n}.
\eea
Here, $\tau_i$ and $\sigma_i$ ($i=x,y,z,0$) are Pauli matrices acting on the crystal sublattice and spin degrees of freedom, respectively. The first and second terms, proportional to $t$ and $t'$, describe nearest and next-nearest-neighbor hopping. The third term represents next-nearest-neighbor spin–orbit coupling with strength $\lambda$. The last term corresponds to the antiferromagnetic exchange interaction of strength $J_n$, where $\hat{\bm n}$ denotes the N\'eel vector direction. Following Ref.~\cite{Jungwirth_prl2017}, we use $t=1$ eV, $t'=0.08$ eV, $\lambda=0.8$ eV, and $J_n=0.6$ eV, with $\hat{\bm n}=\{\cos{\phi},\sin{\phi},0\}$, where $\phi$ is the angle measured from the [100] crystal axis.
\begin{figure}[tb]
    \centering
    \includegraphics[width=1\linewidth]{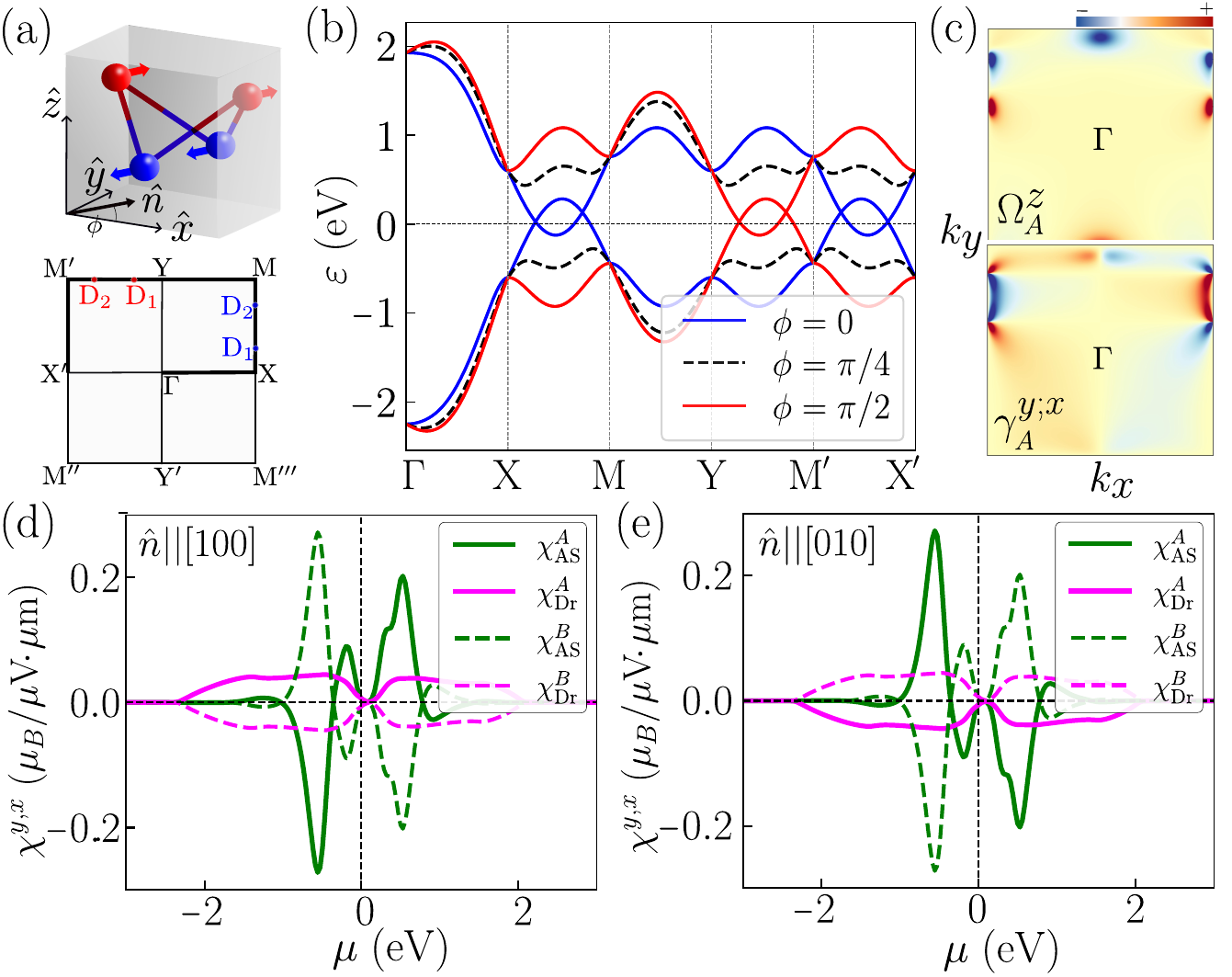}
    \caption{\textbf{N\'eel spin polarization in $\cal PT$-symmetric \ch{CuMnAs}.} 
    (a) Orientation of the N\'eel vector ($\hat{\bm n}$) in quasi-2D tetragonal CuMnAs. Red and blue arrows denote the magnetization directions of Mn atoms on opposite magnetic sublattices. The corresponding two-dimensional Brillouin zone is shown below. 
    (b) Electronic band structure of \ch{CuMnAs} along the path $\Gamma$–X–M–Y–M$'$–X$'$ for different in-plane orientations of the N\'eel order. 
    (c) $\bm k$-space distribution of sublattice-projected Berry curvature (BC) and anomalous spin polarizability (ASP) in the two-dimensional Brillouin zone for $\hat{\bm n}\parallel[100]$. 
    (d),(e) Variation of sublattice-projected spin susceptibility components $\chi^{y;x}$ and $\chi^{x;y}$ with chemical potential for two orthogonal orientations of the N\'eel vector ($\hat{\bm n}\parallel[100]$ and $\hat{\bm n}\parallel[010]$).}
    \label{fig3}
\end{figure}
Diagonalization of Eq.~\eqref{Ham_cumnas} yields two doubly degenerate bands as a consequence of the underlying $\cal PT$ symmetry [see Fig.~\ref{fig3}(a--b)]. For $\phi = 0$ and $\phi = \pi/2$, the system is semimetallic, exhibiting Dirac points ($\mathrm{D}_1$ and $\mathrm{D}_2$) along the high-symmetry paths X–M and Y–$\mathrm{M}'$, respectively. For generic orientations of the N\'eel vector (e.g., $\hat{\bm n}\parallel [110]$, i.e., $\phi=\pi/4$), a full gap opens in the spectrum as shown in Fig.~\ref{fig3}(b). Fig.~\ref{fig3}(c) shows the $\bm{k}$-space distribution of relevant band geometric quantities ($\Omega^{z}$ and $\gamma^{y;x}$) for the valence band projected on sublattice $A$. Fig.~\ref{fig3}(d-e) display the variation of the sublattice-resolved $\cal PT$-odd spin susceptibilities $\chi^{y;x}$ and $\chi^{x;y}$ as a function of the chemical potential $\mu$ for two orthogonal orientations of the N\'eel vector ($\phi=0$ and $\phi=\pi/2$). Notably, the susceptibilities on the two magnetic sublattices exhibit opposite signs, consistent with the $\cal PT$-odd character of the response, as indicated by the solid and dashed curves.

In the above numerical calculations, we have used a momentum relaxation time $\tau=0.1~\mathrm{ps}$, impurity density $n_i\sim 10^{10}~\mathrm{cm}^{-2}$, and first moment of the impurity potential $V_1\sim 0.83\times 10^{-13}~\mathrm{eV\cdot cm}^2$, consistent with parameters employed in Refs.~\cite{Liang_prb2021,Song_prl2023,Fu_SA2020}. For this illustrative parameter set, the anomalous skew-scattering contribution becomes the leading ${\cal P}{\cal T}$-odd contribution over part of the chemical-potential window shown in Fig.~\ref{fig3}. 
For the two-dimensional disorder model, the first moment of the impurity potential $V_1$ carries units of energy times area; this convention is used throughout the numerical estimates. Furthermore, the induced spin polarization on each sublattice is perpendicular to the corresponding local magnetic moment ($\delta \mathbf{S}_{A/B}\perp \mathbf{M}_{A/B}$), thereby generating staggered effective fields that are efficient for electrical switching of the antiferromagnetic order.

\section{Relative impurity scaling of anomalous skew-scattering contribution}
To understand the dependence of the anomalous skew-scattering contribution on impurity parameters, we analyze the scaling of the corresponding spin susceptibility. For this purpose, it is useful to express the microscopic quantities entering the spin response in dimensionless form. We introduce the characteristic energy scale $E_s$ ($\sim 1~\rm eV$) and length scale $a$ (lattice constant) of the system, and define dimensionless variables via following relations,
\be
k_a = \tilde{k}_a/a,~~s^\nu_{mn} = \hbar \tilde{s}^\nu_{mn},~~v^b_{mn} = (E_s a/\hbar)\tilde{v}^b_{mn}. 
\ee

Using these scalings in Eq.~\eqref{New_Spin}, we find that the anomalous skew-scattering susceptibility can be written as
\be
\chi^{\nu;b}_{\rm AS} \sim \frac{e\tau}{\hbar} \, \frac{n_i V_1^3\hbar}{E_s^3 a^6} \, \tilde{\chi}^{\nu;b}_{\rm AS},
\ee
where $\tilde{\chi}^{\nu;b}_{\rm AS}$ is a dimensionless function determined by the intrinsic electronic properties and symmetries of the system. In contrast, the Drude contribution scales as
\be
\chi^{\nu;b}_{\rm Dr} \sim \frac{e\tau}{\hbar} \dfrac{a\hbar}{a^3} \tilde{\chi}^{\nu;b}_{\rm Dr}.
\ee
Therefore, the ratio of the two contributions can be expressed as
\be
\frac{\chi^{\nu;b}_{\rm AS}}{\chi^{\nu;b}_{\rm Dr}} 
= N_{\rm dis} \frac{\tilde{\chi}^{\nu;b}_{\rm AS}}{\tilde{\chi}^{\nu;b}_{\rm Dr}},
\ee
where the dimensionless prefactor $N_{\rm dis} = {n_i V_1^3}/{E_s^3 a^4}$ encodes the dependence on impurity density and potential strength. With the two-dimensional convention used here, $n_i$ carries units of ${\rm cm}^{-2}$, $V_1$ carries units of ${\rm eV\,cm^2}$, $E_s$ carries units of ${\rm eV}$, and $a$ carries units of ${\rm cm}$. The combination $N_{\rm dis} = n_i V_1^3/(E_s^3 a^4)$ is therefore dimensionless, as required for the ratio above. 

\begin{figure}[tb]
    \centering
    \includegraphics[width=0.7\linewidth]{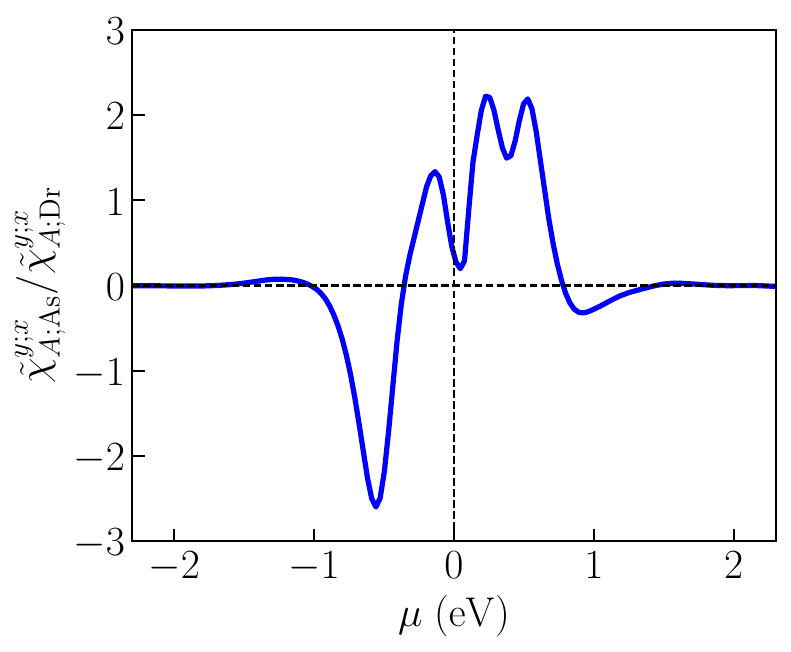}
    \caption{\textbf{Relative strength of dimensionless $\chi^{\nu;b}_{A;\rm AS}$ and $\chi^{\nu;b}_{A;\rm Dr}$}. Variation of the dimensionless sublattice-resolved susceptibility ratio $\tilde{\chi}^{\nu;b}_{A;\rm AS}/\tilde{\chi}^{\nu;b}_{A;\rm Dr}$ with chemical potential. The ratio remains of order unity and varies approximately between $-3$ and $3$.}
    \label{fig4}
\end{figure}

We compute the ratio $\tilde{\chi}^{\nu;b}_{\rm AS}/\tilde{\chi}^{\nu;b}_{\rm Dr}$ numerically for tetragonal CuMnAs as a function of chemical potential. As shown in Fig.~\ref{fig4}, this ratio varies within a range of order unity, typically between $-3$ and $3$. For the illustrative CuMnAs estimate, we take $E_s=1~\rm eV$, $a\approx 3.8~\textup{\AA}$, $n_i\sim 10^{10}~\mathrm{cm}^{-2}$, and $V_1\sim 0.83\times 10^{-13}~\mathrm{eV\cdot cm}^2$. These values are consistent with the CuMnAs-scale disorder parameters used in Refs.~\cite{Liang_prb2021,Song_prl2023,Fu_SA2020}. They give $N_{\rm dis}\approx 2.4$, so the `AS' and `Dr' channels are comparable whenever the dimensionless ratio in Fig.~\ref{fig4} is of order unity. In parameter regions where that intrinsic ratio is also of order unity, the `AS' channel exceeds the Drude contribution, whereas smaller $N_{\rm dis}$ or smaller intrinsic ratios lead to a `Dr'-dominated response. Since $N_{\rm dis}$ scales linearly with impurity density and cubically with the impurity potential, moderate disorder tuning moves the system continuously across the crossover between `Dr' and `AS'-dominated regimes.

\section{N\'eel switching dynamics}

\begin{figure*}[t]
\centering \includegraphics[width=0.8\linewidth]{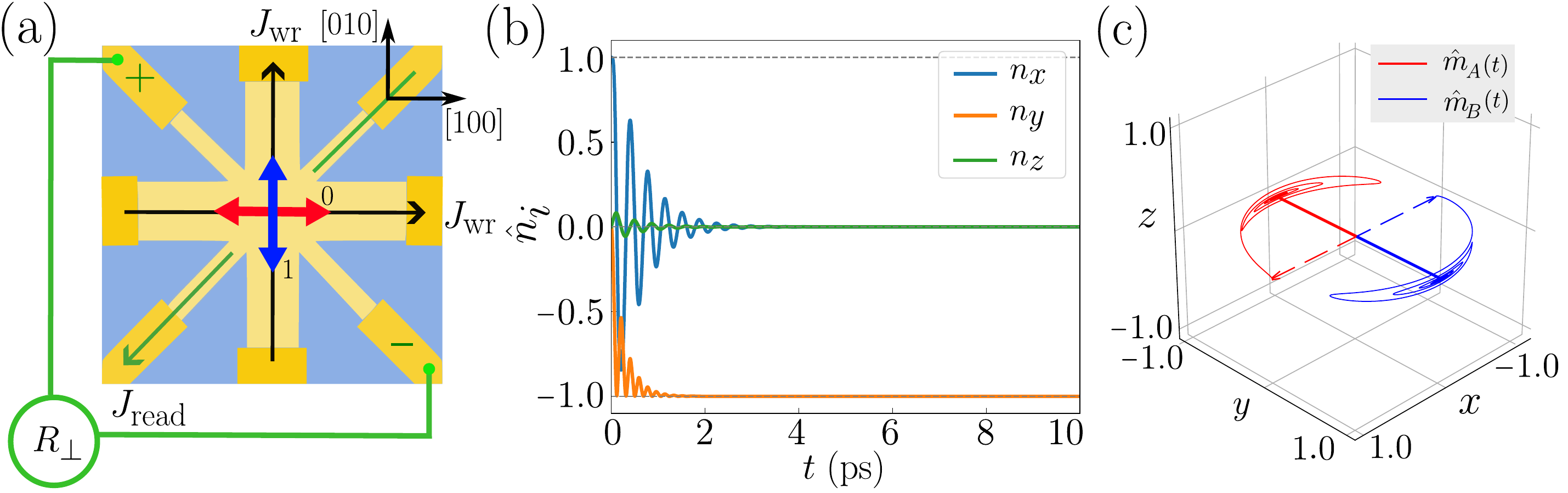}
    \caption{\textbf{N\'eel vector switching in an antiferromagnet.} 
    (a) Schematic of an antiferromagnetic memory device with orthogonal writing terminals for current injection and diagonal probes for readout via anisotropic Hall magnetoresistance. The blue and red double arrows denote two orthogonal orientations of the N\'eel vector, corresponding to the binary memory states (0 and 1). 
    (b) Time evolution of the normalized N\'eel vector components under N\'eel spin–orbit torque (NSOT) driven by asymmetric impurity scattering. The switching between the two states occurs on a timescale of $\sim 5~\rm ps$. 
    (c) Three-dimensional switching trajectory of the N\'eel vector during the reversal process. $\hat{\bm m}_A$ and $\hat{\bm m}_B$ denote unit vectors along the magnetization of the two opposite sublattices.}
\label{fig5}
\end{figure*}

In this section, we demonstrate the impact of NSOT generated by anomalous skew-scattering on the electrical switching of the antiferromagnetic order. A typical antiferromagnetic memory device is illustrated in Fig.~\ref{fig5}(a), where eight terminals are attached to the AFM. The four non-diagonal terminals are used to apply orthogonal bias currents for N\'eel vector switching, while the diagonal terminals are employed to detect the N\'eel vector orientation via anisotropic magnetoresistance measurement. Detailed descriptions of such device architectures and measurement protocols can be found in Refs.~\cite{Wadley_SA2016,Bodnar_NC2018,Tristan_pra2018,Bodner_prb2019,Zhou_pra2018,Godinho_NC2018}. These studies show experimentally that an electric field applied parallel to the N\'eel vector can induce a deterministic $90^\circ$ switching between orthogonal states. Here our aim is to show that the additional anomalous skew-scattering contribution can substantially enhance the torque scale governing such dynamics.

To quantify the role of the anomalous skew-scattering contribution, we analyze the dynamics of the N\'eel vector in tetragonal \ch{CuMnAs}. For a collinear AFM with $A$ and $B$ sublattices, the time evolution of the antiparallel sublattice magnetization unit vectors $\hat{m}_{A,B}$ is governed by the coupled Landau--Lifshitz--Gilbert (LLG) equations~\cite{slonczewski_JOM1996,Liu_prl2011,Yuan_IOP2020,Xu_JAP2023},
\begin{align}\label{Dynamic_AFM}
\frac{d \hat{m}_A}{dt} &= -\gamma (\hat{m}_A \times {\bm B}^{\rm eff}_{A}) + \alpha \left(\hat{m}_A \times \frac{d \hat{m}_A}{dt}\right) + {\bm \tau}^{\rm FL}_{A} + {\bm \tau}^{\rm DL}_{A}, \nn\\
\frac{d \hat{m}_B}{dt} &= -\gamma (\hat{m}_B \times {\bm B}^{\rm eff}_{B}) + \alpha \left(\hat{m}_B \times \frac{d \hat{m}_B}{dt}\right) + {\bm \tau}^{\rm FL}_{B} + {\bm \tau}^{\rm DL}_{B}.
\end{align}
Here, $\gamma$ and $\alpha$ denote the gyromagnetic ratio and Gilbert damping coefficient, respectively. We take $\gamma=1.76\times 10^{11}~\rm s^{-1}T^{-1}$ and, for \ch{CuMnAs}, $\alpha=0.01$~\cite{Selzer_prb2022,Hirst_prb2022,Roy_prb2016}. The effective field acting on each sublattice $i=A,B$ is given by ${\bm B}^{\rm eff}_i = {\bm B}^{\rm ans} + {\bm B}^{\rm ex}_i + {\bm B}^{\rm ext}$, which includes contributions from magnetic anisotropy, inter-sublattice exchange, and any external magnetic field. The exchange field is defined as ${\bm B}^{\rm ex}_{A/B} = -J_{\rm ex}\, {\hat m}_{B/A}/M_s$, where $J_{\rm ex}$ is the exchange coupling strength and $M_s$ is the magnitude of sublattice saturation magnetization. For monolayer tetragonal \ch{CuMnAs}, the exchange field is large, $B^{\rm ex}\sim 700~\rm T$~\cite{Vyborn_prm2020,Wang_prb2020,Belashchenko_apl2018}, while the in-plane anisotropy field is comparatively weak, $B^{\rm ans}\sim 5~\rm mT$~\cite{Vyborn_prm2020,Wang_prb2020,Belashchenko_apl2018}, highlighting the exchange-dominated dynamics of the system.

The field-like (FL) and damping-like (DL) torques originate from the sublattice-resolved local spin polarization $\delta S^{\nu}_{A/B}$. For each sublattice $i$, the spin torques are defined as
${\bm \tau}^{\rm FL}_i = |\tau_i|\, \xi^{\rm FL}\,({\hat m}_i \times \hat{\sigma}^\nu_i)$
and
${\bm \tau}^{\rm DL}_i = |\tau_i|\, \xi^{\rm DL}\, {\hat m}_i \times (\hat{m}_i \times \hat{\sigma}^\nu_i)$,
with the torque magnitude estimated as
$|{\bm \tau}_i| \sim {\gamma J_{\rm ex} |\delta S^{\nu}_i|}/{M_s^2}$~\cite{Cogulu_prl2022,Kondou_NC2021,Zhang_prb2015,Xu_JAP2023}.
Here, $M_s$ is the sublattice saturation magnetization, $\hat{\sigma}^\nu_i$ denotes the spin polarization direction, and $\xi^{\rm FL/DL}$ are the corresponding torque efficiencies. In our calculations, we assume identical efficiencies on both sublattices with $\xi^{\rm DL}/\xi^{\rm FL}=0.25$. 

Using the spin susceptibility obtained from the anomalous skew-scattering mechanism (see Fig.~\ref{fig2}(d--e)), we find a maximum value of $\chi^{y;x}_{\rm AS}\sim-0.25~\mu_B/(\mu{\rm V}\cdot\mu{\rm m})$ at $\mu=-0.7~\rm eV$, which is nearly six times larger than the corresponding Drude contribution. For an applied electric field $|E|=1~\rm V/\mu m$ parallel to the N\'eel vector ($\hat{x}$), the total induced local spin polarization is $\delta S^y=-2\times10^5~\rm\mu_B/\mu m^2$. This results in a torque magnitude $|\tau_i|=1.1\times10^{12}~\rm s^{-1}$, using $M_s\sim27~\mu_B/{\rm nm}^2$~\cite{Wang_prb2020}. The anomalous skew-scattering thus enhances the effective spin torque to approximately six times the conventional contribution.

Substituting these parameters into Eq.~\eqref{Dynamic_AFM} and numerically solving the coupled equations yields the full time evolution of the sublattice magnetizations and the N\'eel vector. As shown in Fig.~\ref{fig5}(b), the system undergoes a coherent $90^\circ$ switching of the N\'eel vector within $\sim 4~\rm ps$. Within the present analysis, the switching trajectory is smooth and deterministic as shown in Fig.~\ref{fig5}(c). This indicates that, for the chosen parameter set, the enhanced NSOT scale is sufficient to drive fast reorientation of the order parameter.

Our LLG calculation is intended as an illustrative dynamics estimate based on the torque scale extracted from the susceptibility calculation. It should not be interpreted as a complete switching phase diagram over damping, pulse shape, and chemical-potential space. Within the chosen parameter set, the asymmetric scattering-induced NSOT produces a substantial enhancement of the spin torque and supports switching on the few-picosecond timescale. An experimentally useful discriminator of this mechanism is the predicted sensitivity of the switching scale to disorder tuning or chemical-potential variation, both of which directly modify the balance between the anomalous skew scattering and Drude channels. This establishes anomalous skew-scattering as a promising and previously unexplored mechanism for efficient electrical control in antiferromagnetic spintronic devices.

\section{Conclusion}

In this work, we have investigated the role of asymmetric impurity scattering in generating staggered current-induced spin polarization in $\cal PT$-symmetric antiferromagnets. Within a semiclassical Boltzmann framework, we identify an additional contribution to the spin susceptibility arising from the interplay between antisymmetric higher-order impurity scattering and the anomalous spin polarizability. We show that this contribution is $\cal PT$-odd and therefore produces a staggered spin polarization on the two magnetic sublattices, enabling a N\'eel spin–orbit torque.

Our analysis establishes a direct connection between this extrinsic mechanism and band-geometric properties of the electronic structure, highlighting the role of band geometry in governing asymmetric scattering processes. Using a minimal model of CuMnAs, we demonstrate that the resulting spin response can be comparable to, and for sufficient magnetic disorder can exceed, the conventional Drude contribution. The corresponding torque is shown to efficiently drive ultrafast switching of the N\'eel vector within a realistic parameter regime.

These results reveal that asymmetric scattering can convert an otherwise $\cal PT$-even response into a staggered spin polarization, providing a new pathway for electrical control of antiferromagnetic order. More broadly, our findings demonstrate that disorder, when combined with band geometry, can play a constructive role in antiferromagnetic spintronics. We expect that this mechanism may be relevant in a wider class of $\cal PT$-symmetric materials and could be explored through controlled disorder engineering in future experiments.

\section{Acknowledgment}

We acknowledge many fruitful discussions with Sunit Das (IIT Kanpur, India), Harsh Varshney (IIT Kanpur, India) and Nirmalya Jana (IIT Kanpur, India). Sayan acknowledges IIT Kanpur for funding support. Amit acknowledges support from the ANRF Core Research Grant CRG/2023/007003, Department of Science and Technology, India.

\section{Data availability}
The data that support the findings of this article are not
publicly available. The data are available from the authors
upon reasonable request.

\onecolumngrid
\appendix
\renewcommand{\theHequation}{app.\Alph{section}.\arabic{equation}}

\section{Spin-shift correction to spin expectation}\label{SS_correction}

In this appendix, we derive the disorder-induced correction to the spin expectation value, $s_{l}^{\nu,{\rm sct}}$, introduced in Eq.~\eqref{total_spin}. This contribution originates from the modification of the Bloch states due to impurity scattering and represents the spin analogue of the side-jump correction. We begin by evaluating the correction to the Bloch state in the presence of an impurity potential $\hat{V}$. Using the Lippmann--Schwinger equation, the exact scattering state is given by
\begin{equation}
\ket{l^{\rm dis}} = \ket{l} + \frac{1}{\varepsilon_l - H_0 + i\eta}\hat{T}\ket{l},
\end{equation}
where the $T$-matrix satisfies $\hat{T} = \hat{V} + \hat{V}(\varepsilon_l - H_0 + i\eta)^{-1}\hat{T}$. The disorder-induced correction to the wave function is then $\ket{\delta^{\rm dis}l} = \ket{l^{\rm dis}} - \ket{l}$. The corresponding correction to the spin expectation value is
\begin{equation}
s_{l}^{\nu,{\rm sct}} = \left\langle 2\,{\rm Re}\bra{l}\hat{s}^{\nu}\ket{\delta^{\rm dis} l} + \bra{\delta^{\rm dis }l}\hat{s}^{\nu}\ket{\delta^{\rm dis}l} \right\rangle_{\rm dis},
\end{equation}
where $\langle \cdots \rangle_{\rm dis}$ denotes disorder averaging. The leading nonvanishing contribution arises at second order in the impurity potential. Expanding $\ket{\delta^{\rm dis}l}$ to second order in $\hat{V}$, we obtain
\begin{align}
\left\langle 2\,{\rm Re}\bra{l}\hat{s}^{\nu}\ket{\delta^{\rm dis} l} \right\rangle_{\rm dis}
&= 2\,{\rm Re} \sum_{l'l''}
\frac{\bra{l}\hat{s}^\nu\ket{l'}\langle V_{l'l''}V_{l''l}\rangle_{\rm dis}}
{(\varepsilon_l-\varepsilon_{l'}+i\eta)(\varepsilon_l-\varepsilon_{l''}+i\eta)}, \\
\left\langle \bra{\delta^{\rm dis }l}\hat{s}^{\nu}\ket{\delta^{\rm dis}l} \right\rangle_{\rm dis}
&= \sum_{l'l''}
\frac{\bra{l'}\hat{s}^\nu\ket{l''}\langle V_{ll'}V_{l''l}\rangle_{\rm dis}}
{(\varepsilon_l-\varepsilon_{l'}-i\eta)(\varepsilon_l-\varepsilon_{l''}+i\eta)}.
\end{align}

Expressing the states in the Bloch basis $|l\rangle = |n\bm{k}\rangle$, we write the impurity matrix elements as $V_{ll'} = \langle n\bm{k}|V|n'\bm{k}'\rangle$. To proceed, we consider a standard model of short-range disorder consisting of randomly distributed $\delta$-function scatterers,
\begin{equation}
V_{\rm imp}(\bm{r}) = \sum_i V_i \delta(\bm{r}-\bm{R}_i).
\end{equation}
Within this model, the matrix element between Bloch states becomes $\bra{n\bm{k}}V\ket{n'\bm{k}'} = V^0_{\bm{k}\bm{k}'} \langle u_{n\bm{k}}|u_{n'\bm{k}'}\rangle,$ where $V^0_{\bm{k}\bm{k}'} = \sum_i V_i e^{i(\bm{k}'-\bm{k})\cdot \bm{R}_i}$. Similarly, the spin operator matrix element reads
\begin{equation}
\bra{n\bm{k}}\hat{s}^\nu\ket{n'\bm{k}'} = \langle u_{n\bm{k}}|\hat{s}^\nu|u_{n'\bm{k}}\rangle\,\delta(\bm{k}'-\bm{k}).
\end{equation}

Substituting these expressions and performing straightforward algebra, we obtain the disorder-induced spin correction in the form
\bea 
s_{l}^{\nu,{\rm sct}}&=&2 {\rm Re} \sum_{\kb'\kb''}\langle V^0_{\kb'\kb''} V^0_{\kb''\kb}\rangle_{\rm dis} \sum_{ n' n''}\dfrac{\langle u_{ n \bm{k}}|\hat{s}^\nu|u_{ n' \bm{k}} \rangle \delta(\kb'-\bm{k})\langle u_{ n'\bm k'}|u_{ n''\bm k''}\rangle \langle u_{ n''\bm k''}|u_{ n\bm k}\rangle}{(\varepsilon_{ n \bm k}-\varepsilon_{ n'\bm k'}+i\eta)(\varepsilon_{ n \bm k}-\varepsilon_{ n''\bm k''}+i\eta)}\nn\\ 
&&~~~~~~~~~~~~~~~~~~~~~~~~~~~~~~~~~~~+\sum_{\kb'\kb''}\langle V^0_{\kb\kb'} V^0_{\kb''\kb}\rangle_{\rm dis}\sum_{ n' n''}\dfrac{ \langle u_{ n' \kb'}|\hat{s}^\nu|u_{ n'' \kb'} \rangle \delta(\kb''-\kb')\langle u_{ n\bm k}|u_{ n'\bm k'}\rangle \langle u_{ n''\bm k''}|u_{ n\bm k}\rangle}{(\varepsilon_{ n \bm k}-\varepsilon_{ n'\bm k'}-i\eta)(\varepsilon_{ n \bm k}-\varepsilon_{ n''\bm k''}+i\eta)}\nn\\ &=& 2{\rm Re} \sum_{\kb''}\langle V^0_{\kb\kb''} V^0_{\kb''\kb}\rangle_{\rm dis} \sum_{ n' n''}\dfrac{\langle u_{ n \bm{k}}|\hat{s}^\nu|u_{ n' \bm{k}} \rangle \langle u_{ n'\bm k}|u_{ n''\bm k''}\rangle \langle u_{ n''\bm k''}|u_{ n\bm k}\rangle}{(\varepsilon_{ n \bm k}-\varepsilon_{ n'\bm k}+i\eta)(\varepsilon_{ n \bm k}-\varepsilon_{ n''\bm k''}+i\eta)}\nn\\ 
&&~~~~~~~~~~~~~~~~~~~~~~~~~~~~~~~~~~~~~~~~~~~~~~~+\sum_{\kb''}\langle V^0_{\kb\kb''} V^0_{\kb''\kb}\rangle_{\rm dis}\sum_{ n' n''}\dfrac{ \langle u_{ n' \kb''}|\hat{s}^\nu|u_{ n'' \kb''} \rangle \langle u_{ n\bm k}|u_{ n'\bm k''}\rangle \langle u_{ n''\bm k''}|u_{ n\bm k}\rangle}{(\varepsilon_{ n \bm k}-\varepsilon_{ n'\bm k''}-i\eta)(\varepsilon_{ n \bm k}-\varepsilon_{ n''\bm k''}+i\eta)}\nn
\eea
\begin{align}\label{sj_correct}
& =2{\rm Re} \sum_{\kb''}\langle V^0_{\kb\kb''} V^0_{\kb''\kb}\rangle_{\rm dis} \sum_{ n' n''}\dfrac{s^\nu_{n n'}(\bm k) \langle u_{ n'\bm k}|u_{ n''\bm k''}\rangle \langle u_{ n''\bm k''}|u_{ n\bm k}\rangle}{(\varepsilon_{ n \bm k}-\varepsilon_{ n'\bm k})(\varepsilon_{ n \bm k}-\varepsilon_{ n''\bm k''}+i\eta)}\notag\\
&\quad +\sum_{\kb''}\langle V^0_{\kb\kb''} V^0_{\kb''\kb}\rangle_{\rm dis}\sum_{ n' n''}\dfrac{s^\nu_{n' n''}(\bm k'') \langle u_{ n\bm k}|u_{ n'\bm k''}\rangle \langle u_{ n''\bm k''}|u_{ n\bm k}\rangle}{(\varepsilon_{ n'' \bm k''}-\varepsilon_{ n'\bm k''})}\left[\dfrac{1}{(\varepsilon_{ n\bm k}-\varepsilon_{ n''\bm k''}+i\eta)}-\dfrac{1}{(\varepsilon_{ n\bm k}-\varepsilon_{ n'\bm k''}-i\eta)}\right]\notag\\
&= 2{\rm Re}\left[\sum_{\kb'}\langle V^0_{\kb\kb'} V^0_{\kb'\kb}\rangle_{\rm dis} \sum_{ n' n''}\left(\dfrac{s^\nu_{n n'}(\bm k) \langle u_{ n'\bm k}|u_{ n''\bm k'}\rangle \langle u_{ n''\bm k'}|u_{ n\bm k}\rangle}{(\varepsilon_{ n \bm k}-\varepsilon_{ n'\bm k})(\varepsilon_{ n \bm k}-\varepsilon_{ n''\bm k'}+i\eta)}\right.+ \left.\dfrac{s^\nu_{n' n''}(\bm k') \langle u_{ n\bm k}|u_{ n'\bm k'}\rangle \langle u_{ n''\bm k'}|u_{ n\bm k}\rangle}{(\varepsilon_{ n'' \bm k'}-\varepsilon_{ n'\bm k'})(\varepsilon_{ n\bm k}-\varepsilon_{ n'\bm k'}-i\eta)}\right)\right]\notag\\
&= 2\pi{\rm Im}\sum_{ n'n''}\sum_{ \kb'}\langle V^0_{\kb\kb'} V^0_{\kb'\kb}\rangle_{\rm dis} \left[\dfrac{s^\nu_{n n''}(\bm k) \langle u_{ n''\bm k}|u_{ n'\bm k'}\rangle \langle u_{ n'\bm k'}|u_{ n\bm k}\rangle}{(\varepsilon_{ n \bm k}-\varepsilon_{ n''\bm k})}-\dfrac{s^\nu_{n' n''}(\bm k') \langle u_{ n\bm k}|u_{ n'\bm k'}\rangle \langle u_{ n''\bm k'}|u_{ n\bm k}\rangle}{(\varepsilon_{ n' \bm k'}-\varepsilon_{ n'' \bm k'})}\right]\delta(\epsilon_{ n \bm{k}}-\epsilon_{ n' \kb'}) \notag\\
&= -2\pi {\rm Im}\sum_{n' n''}\sum_{ \kb'}W_{\kb\kb'}\left[\dfrac{s^\nu_{n' n''}(\bm k') \langle u_{ n\bm k}|u_{ n'\bm k'}\rangle \langle u_{ n''\bm k'}|u_{ n\bm k}\rangle}{(\varepsilon_{ n' \bm k'}-\varepsilon_{ n'' \bm k'})} -\dfrac{s^\nu_{n n''}(\bm k) \langle u_{ n''\bm k}|u_{ n'\bm k'}\rangle \langle u_{ n'\bm k'}|u_{ n\bm k}\rangle}{(\varepsilon_{ n \bm k}-\varepsilon_{ n''\bm k})}\right]\delta(\epsilon_{ n \bm{k}}-\epsilon_{ n' \kb'}).
\end{align}
where we have used $\mathrm{Im}(x+i\eta)^{-1} = -\pi \delta(x)$ and defined the disorder-averaged scattering probability $W_{\bm{k}\bm{k}'} = \langle |V^0_{\bm{k}\bm{k}'}|^2 \rangle_{\rm dis}$. To simplify further, we consider the long-range scattering limit $\bm{q}=\bm{k}'-\bm{k}\to 0$, which is enforced by the intra-band scattering condition ($n'=n$) in Eq.~\eqref{sj_correct}. Expanding the Bloch overlaps to leading order in $\bm{q}$, we obtain
\begin{align}
\langle u_{n\bm{k}}|u_{n\bm{k}'}\rangle \simeq 1 - i q_b \mathcal{R}^b_{nn}~,~~~~~
\langle u_{n\bm{k}}|u_{n''\bm{k}'}\rangle \simeq - i q_c \mathcal{R}^c_{n n''}~,~~~~~
\langle u_{n''\bm{k}}|u_{n\bm{k}'}\rangle \simeq - i q_c \mathcal{R}^c_{n'' n}~.
\end{align}
Substituting these into the above expression and retaining leading-order terms, we obtain
\begin{equation}
s_{l}^{\nu,{\rm sct}}
= -4\pi {\rm Im} \sum_{\bm{k}'} W_{\bm{k}\bm{k}'} 
\delta(\varepsilon_{n\bm{k}}-\varepsilon_{n\bm{k}'})
\sum_{n''\neq n}
\frac{s^\nu_{n n''}(\bm{k})\, i q_b \mathcal{R}^b_{n'' n}}
{\varepsilon_{n\bm{k}}-\varepsilon_{n''\bm{k}}}.
\end{equation}
Using the relation between Berry connection and velocity matrix elements, $v^b_{n''n}(\bm{k}) = \frac{i}{\hbar}(\varepsilon_{n''\bm{k}}-\varepsilon_{n\bm{k}})\mathcal{R}^b_{n''n}$, and the definition of the anomalous spin polarizability $\gamma^{\nu,b}_n(\bm{k})$ as given in Eq.~\eqref{eq:SBC}, we arrive at
\begin{equation}
s_{l}^{\nu,{\rm sct}} = \gamma^{\nu,b}_n(\bm{k}) \sum_{\bm{k}'} w^{(2)}_{n,\bm{k}\bm{k}'} (k_b - k'_b),
\end{equation}
where $w^{(2)}_{n,\bm{k}\bm{k}'} = (2\pi/\hbar) W_{\bm{k}\bm{k}'} \delta(\varepsilon_{n\bm{k}}-\varepsilon_{n\bm{k}'})$ is the leading order Born scattering rate. Finally, expressing the momentum sum in terms of the relaxation time $\tau$, we obtain the compact result~\cite{Guo_prb2024}, 
\begin{equation}
s_{l}^{\nu,{\rm sct}} = \dfrac{k_b\gamma^{\nu,b}_n(\bm{k})}{\tau}.
\end{equation}

\section{Impurity correction to the distribution function}\label{Sct_cor_func}

In this appendix, we solve the semiclassical Boltzmann transport equation to obtain the individual disorder-induced contributions to the non-equilibrium distribution function introduced in Eq.~\eqref{fermi_cont}. The Boltzmann transport equation for the distribution function $f_l$ ($l \equiv (n,\bm{k})$) reads~\cite{Du_NC2019,Guo_prb2024,varshney_2026,sanjay_ESSC2026}
\begin{equation}\label{BTE}
\dot{\bm{k}}\cdot \frac{\partial f_l}{\partial \bm{k}} + \frac{\partial f_l}{\partial t}
= I_{\rm el}(f_l),
\end{equation}
where $I_{\rm el}\{f_l\}$ denotes the collision integral for elastic scattering from static impurities. The collision integral can be written as~\cite{Sinitsyn_jpcm2008,Nagaosa_rmp2010}
\begin{equation}\label{col_int}
I_{\text{el}}(f_l)
= -\sum_{l'} \left( w_{l'l} f_l - w_{ll'} f_{l'} \right),
\end{equation}
where $w_{ll'}$ is the transition rate from state $l$ to $l'$. Within Fermi’s golden rule, it is given by~\cite{ashcroft_book1976_solid}
\begin{equation}
w_{ll'} = \frac{2\pi}{\hbar}
\Big\langle 
\big| \langle l | V_{\text{imp}} | {l'}^{\rm dis} \rangle \big|^{2}
\Big\rangle_{\text{dis}}
\delta(\varepsilon_l - \varepsilon_{l'}).
\end{equation}
Here $V_{\mathrm{imp}}$ is the impurity potential and $|l^{\rm dis}\rangle$ denotes the exact scattering state of the full Hamiltonian $H = H_0 + V_{\mathrm{imp}}$, obtained from the Lippmann--Schwinger equation~\cite{sakurai2017QM}
\begin{equation}\label{eq:LippmannSchwinger}
|l^{\rm dis}\rangle = |l\rangle
+\frac{V_{\text{imp}}}{\varepsilon_l - H_0 + i\eta} |l^{\rm dis}\rangle,
\end{equation}
with $i\eta \to 0^+$ enforcing outgoing boundary conditions. In general, disorder averaging and higher-order scattering processes allow for an antisymmetric component of the transition rate, such that $w_{ll'} \neq w_{l'l}$. It is therefore convenient to decompose
\begin{equation}
w_{ll'}^{s} = \frac{w_{ll'} + w_{l'l}}{2}, \qquad
w_{ll'}^{a} = \frac{w_{ll'} - w_{l'l}}{2},
\end{equation}
where $w^{s}_{ll'} = w^{s}_{l'l}$ and $w^{a}_{ll'} = -w^{a}_{l'l}$. The symmetric part governs conventional relaxation processes, while the antisymmetric part gives rise to skew-scattering effects. For weak disorder, the scattering rate can be expanded perturbatively in powers of the impurity potential. Retaining contributions up to third order~\cite{Xiao_prb2019}, we write
\begin{equation}\label{w_order}
w_{l'l} \simeq w_{l'l}^{(2)} + w_{l'l}^{(3),a}.
\end{equation}
The second-order term is completely symmetric under $l \leftrightarrow l'$ and can be further decomposed into normal and coordinate-shift contributions (discussed in Appendix~\ref{Sct_rate}). The antisymmetric part arises at third order and is responsible for skew scattering. Symmetric higher-order corrections merely renormalize $w^{(2)}$ and are neglected here.

In the presence of a uniform electric field, $\dot{\bm{k}} = -e\bm{E}/\hbar$. Restricting to the steady state ($\partial_t f_l = 0$), Eq.~\eqref{BTE} leads to a hierarchy of coupled equations for the different contributions to the distribution function,
\begin{subequations}\label{BZ_coupl_eq:2}
\begin{align}
-\frac{e\bm E}{\hbar}\cdot \partial_{\bm k} f_l^{\rm in}
&= I_{\rm el}^{\rm in}(f_l^{\rm in}), \\
-\frac{e\bm E}{\hbar}\cdot \partial_{\bm k} f_l^{\rm sj}
&= I_{\rm el}^{\rm in}(f_l^{\rm sj}) + I_{\rm el}^{\rm sj}(f_l^{\rm in}), \\
-\frac{e\bm E}{\hbar}\cdot \partial_{\bm k} f_l^{\rm sk3}
&= I_{\rm el}^{\rm in}(f_l^{\rm sk3}) + I_{\rm el}^{\rm sk3}(f_l^{\rm in}).
\end{align}
\end{subequations}
Here depending on the different physical origins of the impurity scattering rates, the elastic collision integral has been decomposed into distinct contributions. In particular, we separate the collision term into intrinsic, side-jump, and skew-scattering components, $I_{\mathrm{el}}(f_l) = I_{\mathrm{el}}^{\mathrm{in}}(f_l) + I_{\mathrm{el}}^{\mathrm{sj}}(f_l) + I_{\mathrm{el}}^{\mathrm{sk3}}(f_l),$
and these distinct components have the form~\cite{Du_NC2019,varshney_2026,sanjay_ESSC2026},
\begin{align}
\label{eq:collision_terms}
I_{\mathrm{el}}^{\mathrm{in}}(f_l) = - \sum_{l'} w_{ll'}^{\mathrm{2S}} (f_l - f_{l'})~,~~~~~
I_{\mathrm{el}}^{\mathrm{sj}}(f_l) = -\sum_{l'} w_{ll'}^{(2),\mathrm{cs}} (f_l - f_{l'})~,~~~~~~
I_{\mathrm{el}}^{\mathrm{sk3}}(f_l) = \sum_{l'} w_{ll'}^{(3),a} (f_l + f_{l'})~.
\end{align}
The intrinsic collision integral is treated within the relaxation-time approximation,
\be
I_{\rm el}^{\rm in}(f_l^{\rm in})=-\frac{f_l^{\rm in} - f_l^0}{\tau},
\ee
where $\tau$ is the elastic scattering time. We solve these equations perturbatively by expanding the distribution function in powers of the electric field, $f_l = f_l^{(0)} + f_l^{(1)} + f_l^{(2)} + \cdots,$ with $f_l^{(i)} \propto |\bm{E}|^i$. The intrinsic linear-response solution is
\begin{equation}
f_l^{\rm in,(1)} = \frac{e\tau}{\hbar}\,\bm{E}\cdot \partial_{\bm{k}} f_l^0.
\end{equation}
Solving Eq.~\eqref{BZ_coupl_eq:2} to linear order in $\bm{E}$, the side-jump contribution is obtained as
\begin{equation}
f_l^{\rm sj,(1)} = -\tau \sum_{l'} w^{(2),{\rm cs}}_{l'l} \bigl(f_l^0 - f_{l'}^0\bigr).
\end{equation}
Using the identity
\(
v_l^a\,\partial_{\varepsilon_l}\delta(\varepsilon_l-\varepsilon_{l'})
=
\hbar^{-1}\partial_{k_a}\delta(\varepsilon_l-\varepsilon_{l'}),
\)
and performing integration by parts, this can be recast as
\be\label{sj_vel}
f_l^{\rm sj,(1)} = - e\tau\, \bm{E}\cdot \bm{v}_l^{\rm sj}\frac{\partial f^0}{\partial \varepsilon_l},
\ee
where the side-jump velocity is defined as $\bm v_l^{\rm sj}=-\sum_{l'} w^{(2S)}_{l'l}\,\boldsymbol{\delta r}_{ll'},$
with the coordinate shift defined as 
\be\label{shift_def}
\delta {\bm r}_{l'l}=\langle u_{l'} | i\partial_{\bm{k}'} | u_{l'}\rangle-\langle u_l | i\partial_{\bm{k}} | u_l \rangle
-(\partial_{\bm{k}}+\partial_{\bm{k}'})\arg(V_{l'l}). 
\ee
Proceeding similarly, the antisymmetric third-order scattering process yield the skew-scattering corrections
\begin{align}
f_l^{\rm sk3,(1)}&=-\frac{e\tau^2}{\hbar}\sum_{l'} w^{(3),a}_{l'l}\left(\bm{E}\cdot\partial_{\bm k} f_l^0+\bm{E}\cdot\partial_{\bm k} f_{l'}^0
\right).
\end{align}
These expressions provide the linear-order non-equilibrium distribution functions used in the evaluation of the extrinsic spin response discussed in the main text.

\section{Asymmetric scattering rate}\label{Sct_rate}
In this section, we derive simplified expressions for the impurity scattering rates entering the semiclassical Boltzmann equation. Our goal is to reduce the general scattering amplitudes to the forms suitable for numerical evaluation of the spin susceptibilities. The total scattering rate can be decomposed in orders of scattering potential
\be
   w_{ll'}\simeq w_{ll'}^{(2)}+w_{ll'}^{(3)}+\cdots,
\ee
where $w_{ll'}^{(\nu)}\propto V^\nu$ represent the $\nu$-th order scattering, for $\nu=2,3...$ The explicit expressions of these scattering rates are given by
\begin{equation}\label{scrt2_cs}
w_{ll'}^{(2)}=\frac{2\pi}{\hbar}  
\left \langle 
V_{ll'}V_{l'l}
\right \rangle_{\text{dis}} 
\delta(\varepsilon_l - \varepsilon_{l'})~,
\end{equation}
\begin{equation}\label{scat3}
w_{ll'}^{(3)}=\frac{2\pi}{\hbar}  
\sum_{l''}\bigg[\frac{\langle 
V_{ll''}V_{l''l'}V_{l'l}
\rangle_{\text{dis}}}{\varepsilon_{l'} - \varepsilon_{l''}+i\eta} +\frac{\langle 
V_{ll'}V_{l'l''}V_{l''l}
\rangle_{\text{dis}}}{\varepsilon_{l'} - \varepsilon_{l''}-i\eta}\bigg]\delta(\varepsilon_l - \varepsilon_{l'})~.
\end{equation}
The second-order scattering rate $w_{ll'}^{(2)}$ is strictly symmetric under exchange of initial and final states, $l\leftrightarrow l'$. On the other hand, $w_{ll'}^{(3)}$ and $w_{ll'}^{(4)}$ do not follow such relations. The symmetric and antisymmetric parts of these scattering rates can therefore be separated as follows:
\bea
    w_{ll'}^{(\nu),s}=\frac{w_{ll'}^{(\nu)}+w_{l'l}^{(\nu)}}{2}~,~~~w_{ll'}^{(\nu),a}=\frac{w_{ll'}^{(\nu)}-w_{l'l}^{(\nu)}}{2}~.
\eea
The symmetric parts of these higher-order terms merely renormalize the leading second-order contribution and do not generate qualitatively new transport phenomena. Consequently, for the third- and fourth-order scattering processes, only their antisymmetric components are relevant for transport. With this separation, the elastic collision integral naturally decomposes into symmetric and antisymmetric parts as $I^s_{\text{el}}\{f_l\} = -\sum_{l'} w_{l'l}^{(2)} (f_l - f_{l'})$ and $I^a_{\text{el}}\{f_l\} = -\sum_{l'} w_{l'l}^{(3),a}(f_l + f_{l'})$, respectively.

When we account for the work done by the electric field as an electron is displaced within the unit cell during the collision, the second-order impurity contribution is modified as
\bea
\label{scrt2}
w_{ll'}^{(2)}=\frac{2\pi}{\hbar}  
\left \langle 
V_{ll'}V_{l'l}
\right \rangle_{\text{dis}} 
\delta(\varepsilon_l - \varepsilon_{l'}+e{\bm E}\cdot \delta {\bm r}_{ll'})=w^{(2S)}_{ll'}+w^{(2),\rm cs}_{ll'},
\eea
where $\delta {\bm r}_{ll'}$ is the coordinate shift defined in Eq.~\eqref{shift_def} in the main text. The symmetric field-independent part governs conventional momentum relaxation ($w^{(2S)}_{ll'}$), the coordinate shift correction term gives the side jump contribution ($w^{(2),\rm cs}_{ll'}$). Our next step is to simplify these scattering rates and express them in forms suitable for numerical evaluation of the spin conductivity. We begin with the field independent part of the second-order scattering rate $w_{ll'}^{(2S)}$,
\bea
w_{ll'}^{(2S)}&=&\frac{2\pi}{\hbar}  
\left \langle 
V_{ll'}V_{l'l}
\right \rangle_{\text{dis}} 
\delta(\varepsilon_l - \varepsilon_{l'})=\frac{2\pi}{\hbar}  
\left \langle 
V_{\kb\kb'}V_{\kb'\kb}
\right \rangle_{\text{dis}} \langle u_{ n \bm k}|u_{ n' \bm k'}\rangle \langle u_{ n' \bm k'}|u_{ n \bm k}\rangle
\delta(\varepsilon_{ n \bm k} - \varepsilon_{ n'\bm k'})~.
\eea
Here we have used the notation $V_{ll'}=\langle l|V|l'\rangle=\langle n \bm{k}|V| n'\kb'\rangle$. To simplify the expressions above, we adopt the simplest model for disorder, consisting of short-range, randomly distributed $\delta$-function scatterers, $V(\bm{r})=\sum_i V_i\delta(\bm{r}-\bm{R}_i)$. Within this model, the impurity matrix element between Bloch states takes the form
\bea
   \bra{ n \bm k}V\ket{ n'\bm k'}&=&\sum_i\int d{\bm r}V_i\delta(\bm{r}-\bm{R}_i) e^{i(\bm{k}'-\bm{k})\cdot \bm{r}} \langle u_{ n \bm k}|u_{ n' \bm k'}\rangle = \sum_i V_i e^{i(\bm{k}'-\bm{k})\cdot \bm{R}_i} \langle u_{ n \bm k}|u_{ n' \bm k'}\rangle= V^0_{\kb\kb'}\langle u_{ n \bm k}|u_{ n' \bm k'}\rangle~.\nn
\eea
and the disorder average can be further simplified to
\begin{equation}
\left \langle 
V_{\kb\kb'}V_{\kb'\kb}\right \rangle_{\text{dis}}=\left\langle\sum_{ij}V_i V_j \exp\{(\kb'-\kb)(R_i-R_j)\}\right\rangle_{\rm dis}=n_iV_0^2~.
\end{equation}
where $n_i$ is the impurity concentration and $V_0$ is the zeroth moment of the impurity potential. Putting these ingredients together, $w^{(2S)}_{ll'}$ finally simplifies to
\be\label{ScRt2}
w_{ll'}^{(2S)}=\frac{2\pi n_i V_0^2}{\hbar} \langle u_{ n \bm k}|u_{ n' \bm k'}\rangle \langle u_{ n' \bm k'}|u_{ n \bm k}\rangle\delta(\varepsilon_{ n \bm k} - \varepsilon_{ n' \bm k'})~.
\ee
The coordinate-shift component of the scattering rate is given by 
\be
w_{l'l}^{(2),\mathrm{cs}} = \frac{2\pi }{\hbar} 
\langle |V_{l'l}|^{2} \rangle_{\rm dis}\,
\frac{\partial \delta(\varepsilon_l - \varepsilon_{l'})}{\partial \varepsilon_l}\,
e{\bm E}\cdot\delta {\bm r}_{l'l}~,
\ee
which captures the coordinate-shift correction induced by the external electric field $\bm{E}$. This term originates from the real-space displacement of the electron wave packet during an impurity scattering event. The coordinate shift $\delta \bm{r}_{l'l}$ of the wave packet is given by~\cite{sinitsyn_PRB2006}
\begin{equation}\label{coordinate}
\delta {\bm r}_{l'l} = 
\langle u_{l'} | i\partial_{\bm{k}'} | u_{l'} \rangle
-\langle u_l | i\partial_{\bm{k}} | u_l \rangle
-(\partial_{\bm{k}}+\partial_{\bm{k}'})\arg(V_{l'l})~.
\end{equation}
Here, the operator ``$\arg$'' denotes the phase (argument) of a complex number. This coordinate shift plays a crucial role in the side-jump contribution to spin transport. We now turn to the antisymmetric third-order scattering rate,
\begin{equation}
w^{(3),a}_{ll'}=\frac{1}{2}(w_{ll'}^{(3)}-w_{l'l}^{(3)}).
\end{equation}
Substituting Eq.~\eqref{scat3} into this definition, we obtain
\begin{align}\label{sct3}
w_{ll'}^{(3),a}&=(\pi/\hbar)  
\sum_{l''}\bigg[\frac{\langle 
V_{ll''}V_{l''l'}V_{l'l}
\rangle_{\text{dis}}}{\varepsilon_{l} - \varepsilon_{l''}+i\eta} 
+\frac{\langle 
V_{ll'}V_{l'l''}V_{l''l}
\rangle_{\text{dis}}}{\varepsilon_{l} - \varepsilon_{l''}-i\eta}
-\frac{\langle 
V_{l'l''}V_{l''l}V_{ll'}
\rangle_{\text{dis}}}{\varepsilon_{l} - \varepsilon_{l''}+i\eta} 
-\frac{\langle 
V_{l'l}V_{ll''}V_{l''l'}
\rangle_{\text{dis}}}{\varepsilon_{l} - \varepsilon_{l''}-i\eta}\bigg]\delta(\varepsilon_l - \varepsilon_{l'})\notag\\
&= (\pi/\hbar) 
\sum_{l''}\bigg[\langle 
V_{ll''}V_{l''l'}V_{l'l}
\rangle_{\rm dis}\left (\frac{1}{\varepsilon_{l} - \varepsilon_{l''}+i\eta}-\frac{1}{\varepsilon_{l} - \varepsilon_{l''}-i\eta} \right)
-cc.\bigg]\delta(\varepsilon_l - \varepsilon_{l'})\notag\\
&=(4\pi^2/\hbar){\rm Im}\sum_{l''}\langle V_{ll''}V_{l''l'}V_{l'l}
\rangle_{\rm dis}\delta(\varepsilon_{l} - \varepsilon_{l'})\delta(\varepsilon_{l} - \varepsilon_{l''})\notag\\
&= (4\pi^2/\hbar){\rm Im}\sum_{n''} \sum_{ \bm k''}\langle V_{\kb\kb''}^0 V_{\kb''\kb'}^0 V_{\kb'\kb}^0
\rangle_{\rm dis}\langle u_{ n \bm k}|u_{ n'' \bm k''} \rangle\langle u_{ n'' \bm k''}|u_{ n' \bm k'}\rangle \langle u_{ n' \bm k'}|u_{ n \bm k}\rangle\delta(\varepsilon_{ n \bm k} - \varepsilon_{ n' \bm k'})\delta(\varepsilon_{ n\bm k} - \varepsilon_{ n'' \bm k''})~.
\end{align}
With the same approximations as before, we first evaluate the disorder-averaged third-order impurity correlator as 
\be\label{sct3_pot}
\langle V_{\kb\kb''}^0 V_{\kb''\kb'}^0 V_{\kb'\kb}^0
\rangle_{\rm dis}=\sum_{i,j,k}V_i V_j V_k \exp\left((\kb''-\bm{k})R_i+(\kb'-\kb'')R_j+(\bm{k}-\kb')R_k\right)=n_iV_1^3~,
\ee
with $n_i$ being the impurity concentration and $V_1$ being the first moment of the impurity potential. Finally the antisymmetric third-order scattering rate (non-Gaussian skew scattering) takes the compact form
\bea
w_{ll'}^{(3),a}=\frac{4\pi^2 n_i V_1^3}{\hbar}{\rm Im}\sum_{\bm k''}\langle u_{ n \bm k}|u_{ n'' \bm k''} \rangle\langle u_{ n'' \bm k''}|u_{ n' \bm k'}\rangle \langle u_{ n' \bm k'}|u_{ n \bm k}\rangle\delta(\varepsilon_{ n \bm k} - \varepsilon_{ n' \bm k'})\delta(\varepsilon_{ n\bm k} - \varepsilon_{ n'' \bm k''})~.
\eea
Within the present set of approximations, this represents the maximal level of simplification. To proceed further, we need to explicitly evaluate the Bloch-state overlaps entering the disorder matrix elements. We note that the Bloch wave functions satisfy the orthogonality relation $\langle  n \bm{k}| n'\kb'\rangle=\delta_{\kb\kb'}\delta_{ n  n'}$, but the periodic part of the Bloch wave function need not satisfy the same relation, i.e. $\langle u_{ n \bm k}|u_{ n' \bm k'}\rangle \neq \delta_{\kb\kb'}\delta_{ n  n'}$. This means there can be finite overlap between $\ket{u_{ n' \bm k'}}$ and $\ket{u_{ n \bm k}}$. 

Here we assume that no inter-band transition occurs during scattering. This imposes $n=n'$ for $w^{(2S)}_{ll'}$ and $n=n'=n''$ for $w^{(3),a}_{ll'}$. This condition effectively simplifies the Bloch-state overlaps, making them finite only in the regime of small momentum transfer $q=(\kb'-\bm{k})\to 0$. Within this limit, we expand the Bloch state $\ket{u_{ n \kb'}}$ around $\bm{k}$ as
\be\label{expand_u}
\ket{u_{ n \kb'}}=\ket{u_{ n \bm{k}}}+q_b \ket{\partial_b u_{ n \bm{k}}}+\frac{1}{2}q_b q_c\ket{\partial_b \partial_c u_{ n \bm{k}}}+\cdots~.
\ee
Here we adopt Einstein summation over the repeated coordinate index $b$. This leads to
\be\label{expand_uu}
\langle u_{ n \bm k}|u_{ n \bm k'}\rangle\approx 1-iq_b\mathcal{R}_{ n n}^{b}(\bm k)\approx e^{-iq_b\mathcal{R}^b_{ n n}(\bm k)}~. 
\ee
Using this in Eq.~\eqref{ScRt2}, we obtain
\be\label{final_SR2}
   w^{(2S)}_{ll'}=w_{n,\kb\kb'}^{(2S)}=\frac{2\pi n_i V_0^2}{\hbar} |\langle u_{ n \bm k}|u_{ n \bm k'}\rangle|^2 \delta(\varepsilon_{ n \bm k} - \varepsilon_{ n \bm k'})\approx \frac{2\pi n_i V_0^2}{\hbar} \delta(\varepsilon_{ n \bm k} - \varepsilon_{ n \bm k'})~.
\ee
We now consider the case of $w_{ll'}^{(3),a}$ within the same limit. It contains $\langle u_{ n \bm k}|u_{ n \bm k'}\rangle$, $\langle u_{ n \kb''}|u_{ n \kb'}\rangle$ and $\langle u_{ n \kb'}|u_{ n \bm k}\rangle$. We also assume $q'=(\kb''-\bm{k})\to0$, which leads to $\ket{u_{ n \kb''}}=\ket{u_{ n \bm{k}}}+q_b' \ket{\partial_b u_{ n \bm{k}}}+\frac{1}{2}q_b' q_c'\ket{\partial_b \partial_c u_{ n \bm{k}}}+\cdots$. The above overlaps then evaluate to
\bea
    \langle u_{ n \bm k}|u_{ n \bm k''}\rangle &=&1-iq_b'\mathcal{R}^b_{ n n}+\frac{1}{2}q_b'q_c'\langle u_{ n \bm k}|\partial_b \partial_c u_{ n \bm k}\rangle~,\nn\\
    \langle u_{ n \bm k''}|u_{ n \bm k'}\rangle &=& 1+i(q_b'-q_b)\mathcal{R}_{ n n}^b +q_b q_c'\langle \partial_c u_{ n \bm{k}}|\partial_b u_{ n \bm{k}}\rangle+\frac{1}{2}q_b q_c \langle u_{ n \bm k}|\partial_b \partial_c u_{ n \bm k}\rangle+\frac{1}{2}q_b' q_c'\langle \partial_b \partial_c u_{ n \bm k}|u_{ n \bm k}\rangle~,\nn\\
    \langle u_{ n \bm k'}|u_{ n \bm k}\rangle &=& 1+iq_b\mathcal{R}^b_{ n n}+\frac{1}{2}q_b q_c\langle \partial_b \partial_c u_{ n \bm k}| u_{ n \bm k}\rangle~. \nn
\eea
Using the above in Eq.~\eqref{sct3}, we can simplify $w_{n,\kb\kb'}^{(3),a}$ as follows
\begin{align}\label{w_3a_final}
    w^{(3),a}_{ll'}=w_{n,\kb\kb'}^{(3),a}&=\frac{4\pi^2 n_i V_1^3}{\hbar}{\rm Im}\sum_{\bm k''}\langle u_{ n \bm k}|u_{ n \bm k''} \rangle\langle u_{ n \bm k''}|u_{ n \bm k'}\rangle \langle u_{ n \bm k'}|u_{ n \bm k}\rangle\delta(\varepsilon_{ n \bm k} - \varepsilon_{ n \bm k'})\delta(\varepsilon_{ n\bm k} - \varepsilon_{ n \bm k''})\notag\\
    &= \frac{4\pi^2 n_i V_1^3}{\hbar}\sum_{\bm k''}{\rm Im}[1+(q_bq_c+q_b'q_c'-q_bq_c')\mathcal{R}^b_{ n n}\mathcal{R}^c_{ n n}+q_bq_c'\langle \partial_c u_{ n\bm{k}}|\partial_b u_{ n\bm{k}}\rangle\notag\\
    &\quad +(q_b q_c+q_b' q_c'){\rm Re}(\langle u_{ n \bm{k}}|\partial_b\partial_c u_{ n \bm{k}}\rangle)]\delta(\varepsilon_{ n \bm k}- \varepsilon_{ n \bm k'})\delta(\varepsilon_{ n\bm k} - \varepsilon_{ n \bm k''})\notag\\
    &= \frac{4\pi^2 n_i V_1^3}{\hbar}\sum_{ \kb''}q_b q_c'{\rm Im}[\langle \partial_c u_{ n\bm{k}}|\partial_b u_{ n\bm{k}}\rangle]\delta(\varepsilon_{ n \bm k}- \varepsilon_{ n \bm k'})\delta(\varepsilon_{ n\bm k} - \varepsilon_{ n \bm k''})\notag\\
    &= -\frac{2\pi^2 n_i V_1^3}{\hbar}\sum_{ \kb''}(k_b'-k_b) (k_c''-k_c)\Omega^{bc}_ n(\bm{k}) \delta(\varepsilon_{ n \bm k}- \varepsilon_{ n \bm k'})\delta(\varepsilon_{ n\bm k} - \varepsilon_{ n \bm k''})\notag\\
    &= \frac{2\pi^2 n_i V_1^3}{\hbar}\sum_{ \kb''}[(k_b'-k_b) (k_c''-k_c)]\epsilon_{bcd}\Omega^{d}_ n(\bm{k}) \delta(\varepsilon_{ n \bm k}- \varepsilon_{ n \bm k'})\delta(\varepsilon_{ n\bm k} - \varepsilon_{ n \bm k''})\notag\\
     &= \frac{2\pi^2 n_i V_1^3}{\hbar}\sum_{ \kb''}[(\kb'-\bm{k})\times(\kb''-\bm{k})]\cdot\bm{\Omega}_ n(\bm{k}) \delta(\varepsilon_{ n \bm k}- \varepsilon_{ n \bm k'})\delta(\varepsilon_{ n\bm k} - \varepsilon_{ n \kb''})\notag\\
      &= -\frac{2\pi^2 n_i V_1^3}{\hbar}\sum_{ \kb''}[(\kb''\times\kb')+ (\kb'\times\bm{k})+(\bm{k}\times\kb'')]\cdot\bm{\Omega}_ n(\bm{k}) \delta(\varepsilon_{ n \bm k}- \varepsilon_{ n \bm k'})\delta(\varepsilon_{ n\bm k} - \varepsilon_{ n \bm k''})~.
\end{align}
Here $\Omega_{ n}^{bc}$ is the Berry curvature of the system and is defined as $\Omega_{ n}^{bc}=-\Omega_{ n}^{cb}=2\mathrm{Im}[\langle\partial_c u_{n\bm k}|\partial_b u_{n\bm k}\rangle]$ for the $n$-th band.

\twocolumngrid
\bibliography{refs}
\end{document}